\documentclass[showpacs,preprintnumbers,12pt,amsmath,amssymb]{revtex4}
\usepackage{amsmath}
\usepackage{graphicx}
\usepackage{booktabs}

\usepackage{color}

\begin{document}

\title{Generation of discrete breathers from linear to nonlinear Anderson disorder in electrical lattices}

\author{Xuan-Lin Chen$^{1}$}
\author{F. Palmero$^{2}$}

\affiliation{$^{1}$Department of Physics, Heilongjiang University, Harbin, Heilongjiang 150000, China}

\affiliation{$^{2}$ Grupo de F\'isica No Lineal, Departamento de F\'isica Aplicada I, ETSI Inform\'atica, Universidad de Sevilla, Avda Reina Mercedes s/n, E-41012 Sevilla, Spain}

\date{\today}

\begin{abstract}
We investigate linear and nonlinear localization in a disordered electrical lattice composed of coupled LC resonators. In the linear regime, the normal-mode spectrum and spatial voltage profiles are characterized theoretically and experimentally, revealing strongly localized Anderson modes induced by capacitive disorder. Good agreement is found between numerical predictions and experimental measurements of resonance frequencies, spatial profiles, inverse participation ratios, and localization lengths. Nonlinearity is introduced through varactor diodes, allowing the numerical continuation of a selected Anderson-localized mode into a family of discrete breathers. Floquet analysis indicates that these nonlinear localized solutions remain spectrally stable over the investigated frequency range, while their spatial localization evolves with increasing amplitude. Experimental measurements reveal nonlinear resonance bending, frequency shifts, and hysteresis, in qualitative agreement with the predicted nonlinear behavior. These results provide numerical evidence for the emergence of stable, disorder-pinned discrete breathers from Anderson-localized modes, supported by experimental observations of the associated nonlinear resonant response, and highlight the interplay between disorder and nonlinearity in electrical lattices.
\end{abstract}


\maketitle

\section{INTRODUCTION}
Since Anderson proposed the concept of localization in 1958, Anderson localization has remained an enduring research topic in condensed matter physics and wave physics, and its universality has been extensively verified in various disordered systems \cite{anderson1958, lagendijk2009, abrahams1979}. In condensed matter electronic systems, studies based on monolayer van der Waals metals \cite{thinel2026}, the two-dimensional random gap model \cite{hill2014}, and three-dimensional disordered models \cite{bhattacharjee2025, devakul2017} have revealed dimensionality-reduction-driven localization and universality class behavior. In cold-atom systems, the exponential scaling relation verified by kicked rotor experiments \cite{manai2015}, the forward-scattering peak as a signature of strong localization in quantum quenches \cite{micklitz2014}, and the destruction of localization by interactions between bright solitons \cite{plodzien2012} have all been reported. In photonics, boundary and nonlinear effects in photonic lattices \cite{jovic2011, jovic2012}, near-visible light localization \cite{riboli2011}, temporal-domain localization \cite{eswaran2025}, competition between dynamic and Anderson localization \cite{borovkova2015}, and coexistence of flat-band and Anderson localization \cite{liu2026} have been successively revealed. In phononic systems, the exponential decay of mode-resolved transmission in Si/Ge superlattices has provided direct evidence for phonon localization \cite{hu2021}. In non-Hermitian and topological systems, skin-Anderson hybrid modes \cite{shang2026} and scale-free localization \cite{yilmaz2024} have further expanded the scope of localization. In terms of theoretical methods, tools such as the localization tensor \cite{tao2022}, the typical medium cluster approximation \cite{ekuma2014}, and the speckle similarity function \cite{leseur2014} have provided support for the study of phase transitions. These continuous cross-platform and cross-dimensional explorations have made Anderson localization an active frontier to this day. In addition, circuit systems, with their flexibly tunable coupling and grounding design, offer another highly controllable artificial platform for simulating and observing various localization phenomena.

Previously, research in circuit systems has been extended to various localization phenomena, mainly focusing on topological localization \cite{english2026}, flat-band localization \cite{chase2024}, scale-tailored localization \cite{guo2024}, linear impurity modes \cite{molina2019}, and nonlinear intrinsic localized modes \cite{english2008, english2010, palmero2011, sato2007, stearrett2007}, etc. In recent years, the study of Anderson localization in circuit systems has also attracted increasing attention. Based on topological circuits, non-Abelian Anderson localization and transitions have been experimentally realized, with delocalization, coexistence, and localization phases observed \cite{wang2023}. The competition and interconversion between the non-Hermitian skin effect and Anderson localization have been experimentally verified in a nonreciprocal Aubry-Andr\'{e} model, where the localization center can be manipulated by tuning the nonreciprocity strength or disorder strength \cite{halder2025, du2025}. The inverse Anderson transition has also been realized in an Aharonov-Bohm circuit through antisymmetric-correlated disorder, where disorder induces a transition from localized states to extended states \cite{wang2022}. These works demonstrate that circuit systems provide an ideal experimental platform for exploring Anderson localization and its interplay with nonlinearity, non-Hermiticity, and topology. However, in circuit-based studies of Anderson localization, a unified theoretical and experimental picture of linear localization and nonlinear discrete breathers (DBs) remains insufficiently explored.

In this paper, we construct a basic Anderson model with both linear and nonlinear electronic components and investigate it theoretically and experimentally. The theoretical and experimental results are in good agreement. We first analyze the resonance frequencies, localization centers, voltage profiles, inverse participation ratio (IPR) values, and localization lengths of the linear model. 
Then we use the nonlinear model to generate the DBs, and analyze their stability, voltage profiles and resonance characteristics.

\section{THE THEORETICAL MODEL}
\begin{figure}[htbp]
    \centering
    \includegraphics[width=0.9\textwidth]{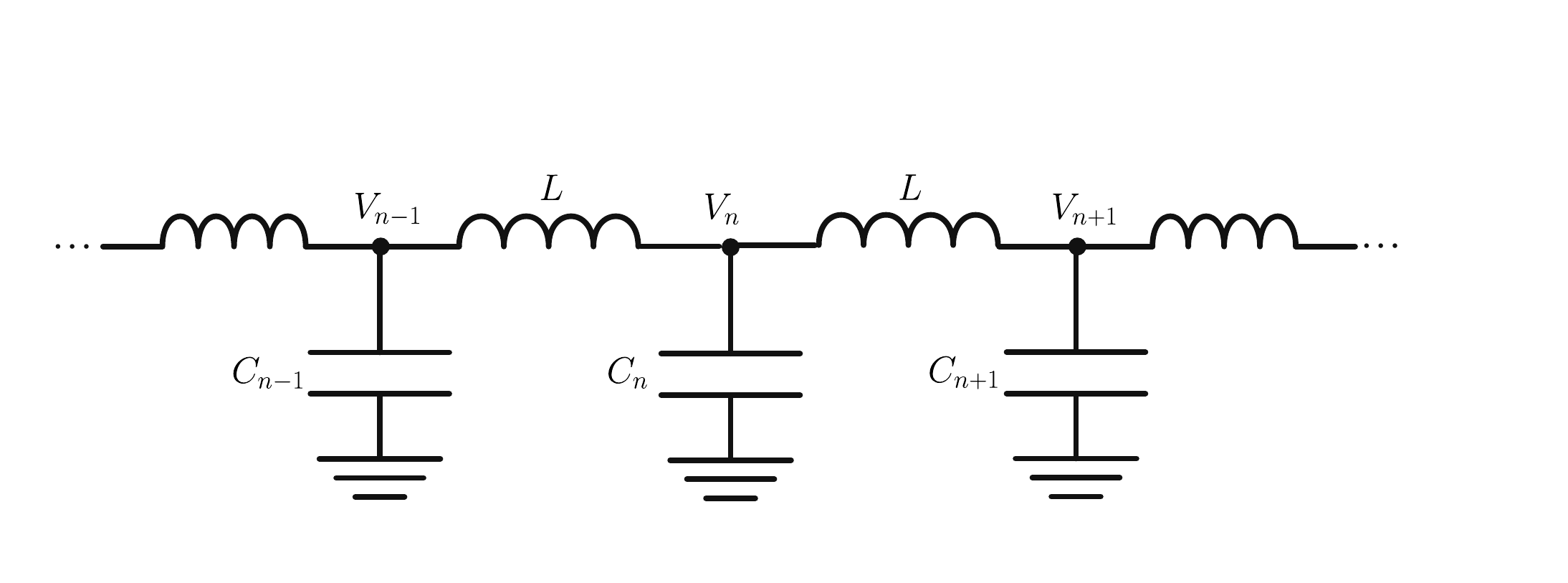}
    \caption{Sketch of the electrical lattice.}
    \label{fig1}
\end{figure}

Our electrical lattice consists of a set of nodes, each connected to ground through a capacitor $C_n$, with neighboring nodes interconnected by inductors $L$, as shown in Fig.~\ref{fig1}. Using Kirchhoff’s laws and assuming ideal circuit elements, namely lossless inductors and linear capacitors, the equations governing the system are
\begin{equation}
	C_n\frac{d^2V_n}{dt^2}=\frac{1}{L}\left(V_{n+1}+V_{n-1}-2V_n\right),
	\label{And_eq_9}
\end{equation}
For a harmonic normal mode $V_n(t)~=~v_n\exp(\mathrm{i}\omega t)$, and separating the uniform part of the capacitance $C_0$ from its variation $(C_n-C_0)$, Eq. (\ref{And_eq_9}) becomes
\begin{equation}
	\epsilon_n(\omega) v_n-(v_{n+1}+v_{n-1})=E(\omega) v_n,
	\label{And_eq_10}
\end{equation}
where $\epsilon_n=(C_0-C_n)L\omega^2$ and $E(\omega)=LC_0\omega^2-2$. 
This equation has the same mathematical structure as the one-dimensional Anderson equation $\epsilon_n c_n-s\left(c_{n+1}+c_{n-1}\right)=E c_n$. Here $\epsilon_n$ denotes the onsite potential, $E$ is the energy eigenvalue, $s$ is the hopping constant, and $c_n$ is the probability amplitude of the $n$th node. The hopping coefficient is normalized to unity in our model.
However, the correspondence does not imply complete physical equivalence, since the effective on-site disorder in the electrical lattice depends explicitly on the frequency.

\begin{figure}[htbp]
    \centering
    \includegraphics[width=0.8\textwidth]{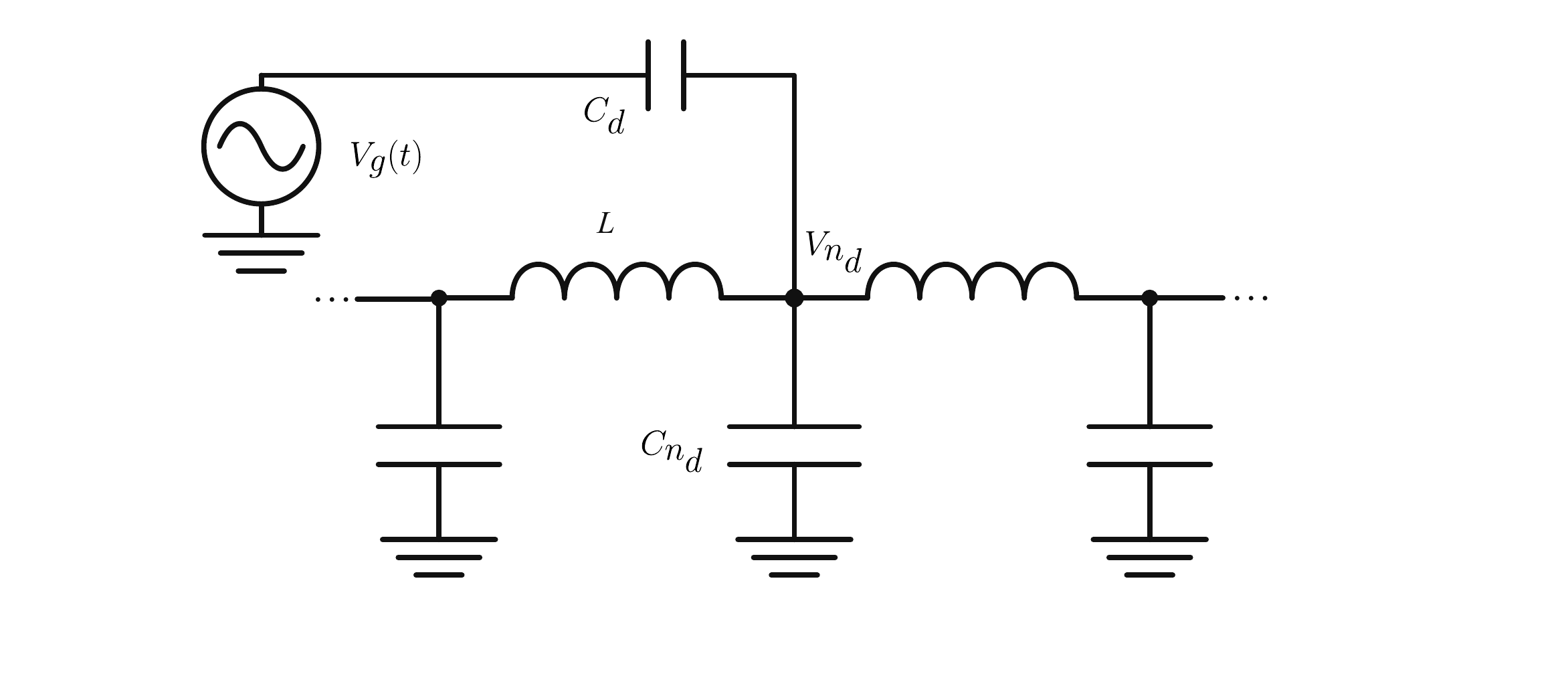}
    \caption{Schematic of the driven node in the electrical lattice.}
    \label{fig2a}
\end{figure}

Figure~\ref{fig2a} shows a schematic of the driven node of the electrical lattice. To excite the normal modes of the lattice without significantly perturbing its dynamics, a sinusoidal voltage $V_{\mathrm{g}}(t)$ was applied to a selected node $n_{\mathrm{d}}$ through a small coupling capacitor $C_{\mathrm{d}}$, with $C_{\mathrm{d}}\ll C_n$. For a harmonic excitation of angular frequency $\omega$, the current injected into the lattice is
\begin{equation}
I_{\mathrm{d}}
=
C_{\mathrm{d}}
\left(\dot{V}_{\mathrm{g}}-\dot{V}_{n_{\mathrm{d}}}\right),
\end{equation}
where an overdot denotes differentiation with respect to time. 

To construct a linear disordered lattice for characterizing Anderson modes, we connect a capacitor $C_n^*$ in parallel with a $C_0=680 \mathrm{pF}$ capacitor at each node $n$. The resulting effective nodal capacitance is therefore $C_n=C_0+C_n^*$. By selecting the values of $C_n^*$ randomly from a prescribed range, a spatially disordered distribution of nodal capacitances is obtained. This capacitive disorder plays the role of an effective random on-site potential and gives rise to Anderson-localized normal modes. These modes can then be characterized by measuring their resonance frequencies and spatial voltage profiles across the lattice. 

\begin{figure}[h]
\centering
\includegraphics[width=0.3\textwidth]{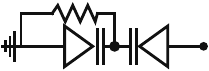}
\caption{Schematic representation of the nonlinear element used to replace each $C_0$ capacitor.}
\label{fig6}
\end{figure}

For studying a nonlinear system without substantially modifying the lattice architecture, we just replaced the $C_0=680$ pF capacitors at the lattice sites with nonlinear elements, each consisting of two NTE618 varactor diodes and a 100 k$\Omega$ resistor, following the configuration described in Ref.~\cite{chase2024} and shown schematically in Fig.~\ref{fig6}.
The capacitance of the element $C^*$is a function of the voltage at each node $V_n$, and it can be approximated as ~\cite{chase2024}
\begin{equation}
    C^*(V_n)=\frac{C_{0}^{*}}{1+\ln(1+g V_n^2)},
    \label{And_eq_13}
\end{equation}
where $C_{0}^{*} \approx 450$ pF and $g=0.1$.
Using Kirchhoff's laws and assuming ideal circuit elements, the equations governing the system are
\begin{equation}
	\left[C_n+C^*(V_n)\right]\frac{d^2V_n}{dt^2}+\frac{d C^*}{d V_n} \left(\frac{d V_n}{d t}\right)^2=\frac{1}{L}\left(V_{n+1}+V_{n-1}-2V_n\right),
	\label{And_eq_14}
\end{equation}
with periodic boundary conditions.

\section{RESULTS AND DISCUSSION}

We consider a chain of $N=21$ nodes with periodic boundary conditions and use the IPR to identify localization,
\begin{equation}
	\mathrm{IPR}
	=
	\frac{\sum_{n=1}^{N}|v_n^{max}|^4}
	{\left(\sum_{n=1}^{N}|v_n^{max}|^2\right)^2}.
    \label{And_eq_11}
\end{equation}
A mode concentrated on only a few nodes has a comparatively large IPR,
whereas an extended mode has an IPR of order $1/N$. In a finite lattice, localization is clearly observable only for modes whose localization length $\xi$ is significantly smaller than the system size. 
The localization length $\xi$ of a spatially localized mode is defined as the characteristic distance over which its amplitude envelope decreases by a factor $e$ away from its localization center. If $n_0$ is the node of maximum amplitude, the asymptotic spatial profile is expected to behave as $|V_n|\propto \exp\left[-\frac{d(n,n_0)}{\xi}\right],$ where, for a ring of $N$ nodes with periodic boundary conditions, $d(n,n_0)=\min\left(|n-n_0|,,N-|n-n_0|\right)$.  Numerically, it can be obtained by fitting $\ln |V_n|$ as a function of $d(n,n_0)$ over the exponentially decaying tails of the mode. If the fitted slope is $s<0$, then $\xi=-1/s$.  If the squared amplitude $|V_n|^2$ is fitted instead, then $\xi=-2/s$. 

\subsection{The linear chain with Anderson disorder}

The first step in exploring Anderson localization in a linear chain is to select an appropriate random capacitance distribution. The disordered $C_n^*$ we choose are listed in Table \ref{tab:additional_capacitors}, they were selected because they are standard commercial values that are inexpensive and readily available. As will also be shown in our subsequent results, their selection involves no additional special considerations. When connected $C_n^*$ in parallel with the base capacitor $C_0$ they produce effective nodal capacitances ranging from 680 to 1680 pF. This range provides sufficiently strong capacitive disorder to make spatial localization observable while keeping the normal-mode frequencies within the operating and measurement range of the experimental equipment. The value $C_n^*=0$ corresponds to nodes at which no additional capacitor was connected. 
\begin{table}[h]
\centering
\caption{Additional capacitors connected in parallel with $C_0$ at each lattice node.}
\label{tab:additional_capacitors}   
\renewcommand{\arraystretch}{1.15}
    \begin{tabular}{cc cc cc}
        \toprule
        \textbf{Node $n$} & \(\boldsymbol{C_n^*}\) \textbf{(pF)}
        & \textbf{Node $n$} & \(\boldsymbol{C_n^*}\) \textbf{(pF)}
        & \textbf{Node $n$} & \(\boldsymbol{C_n^*}\) \textbf{(pF)}
        \\
         1 &    0 &  8 & 1000 & 15 &  220 \\
         2 &    0 &  9 &    0 & 16 &  220 \\
         3 &  470 & 10 & 1000 & 17 & 1000 \\
         4 &  220 & 11 &  680 & 18 & 1000 \\
         5 &  220 & 12 &    0 & 19 &    0 \\
         6 &  220 & 13 &    0 & 20 & 1000 \\
         7 &    0 & 14 &    0 & 21 &  220 \\
    \end{tabular}
\end{table}

Using Eq.~(\ref{And_eq_10}), the determination of the normal modes reduces to an eigenvalue problem that can be solved numerically. The IPR of each mode and its corresponding effective localization length $\xi$ can then be readily calculated from the numerical eigenvectors. The resulting mode frequencies, IPR values, principal localization nodes, and localization lengths are summarized in Table~\ref{tab:exponential_localization_lengths}.

\begin{table}[htbp]
\centering
\caption{Inverse participation ratios and localization lengths obtained
from exponential fits to the two spatial tails of the linear normal modes.}
\label{tab:exponential_localization_lengths}
\renewcommand{\arraystretch}{1.1}
\begin{tabular}{ccccc}
\toprule
\textbf{Mode} &
\(\boldsymbol{f}\) \textbf{(kHz)} &
\textbf{IPR} &
\textbf{Main node} &
\(\boldsymbol{\xi_{\mathrm{exp}}}\) \\
\midrule
 0 &  0.000 & 0.0476 & all    & ---               \\
 1 &  4.370 & 0.0730 & 8      & ---               \\
 2 &  4.973 & 0.0704 & 3      & ---               \\
 3 &  9.068 & 0.0757 & 20     & \(23.21^\dagger\) \\
 4 &  9.430 & 0.0763 & 8      & ---               \\
 5 & 12.917 & 0.0824 & 14     & \(60.19^\dagger\) \\
 6 & 14.235 & 0.0819 & 5      & \(18.12^\dagger\) \\
 7 & 17.059 & 0.0904 & 13     & ---               \\
 8 & 18.348 & 0.0778 & 1      & ---               \\
 9 & 20.115 & 0.0760 & 1      & \(59.86^\dagger\) \\
10 & 22.315 & 0.1139 & 13     & \(9.96^\dagger\)  \\
11 & 23.374 & 0.0932 & 4      & \(6.28^\dagger\)  \\
12 & 27.212 & 0.1107 & 2      & \(16.94^\dagger\) \\
13 & 27.618 & 0.1449 & 16     & ---               \\
14 & 29.617 & 0.1727 & 9      & \(23.33^\dagger\) \\
15 & 30.639 & 0.3437 & 19 (9) & ---               \\
16 & 30.907 & 0.1533 & 9 (19) & ---               \\
17 & 31.726 & 0.2123 & 12     & \(3.42^\dagger\)  \\
18 & 33.514 & 0.3075 & 7      & 1.84              \\
19 & 35.467 & 0.3750 & 1      & 1.14              \\
20 & 36.540 & 0.3095 & 13     & 0.92              \\
\end{tabular}
\medskip
\begin{minipage}{0.96\linewidth}
\footnotesize
The symbol \(\dagger\) indicates that the reported localization length is
only a formal estimate because at least one of the exponential fits has a
low coefficient of determination. A dash denotes a uniform mode, a mode
without a clear two-sided exponential decay, or a mode with more than one
significant localization center. The localization length
\(\xi_{\mathrm{exp}}\) is expressed in lattice sites.
\end{minipage}
\end{table}

\begin{figure}[h]
\centering
\includegraphics[width=\textwidth]{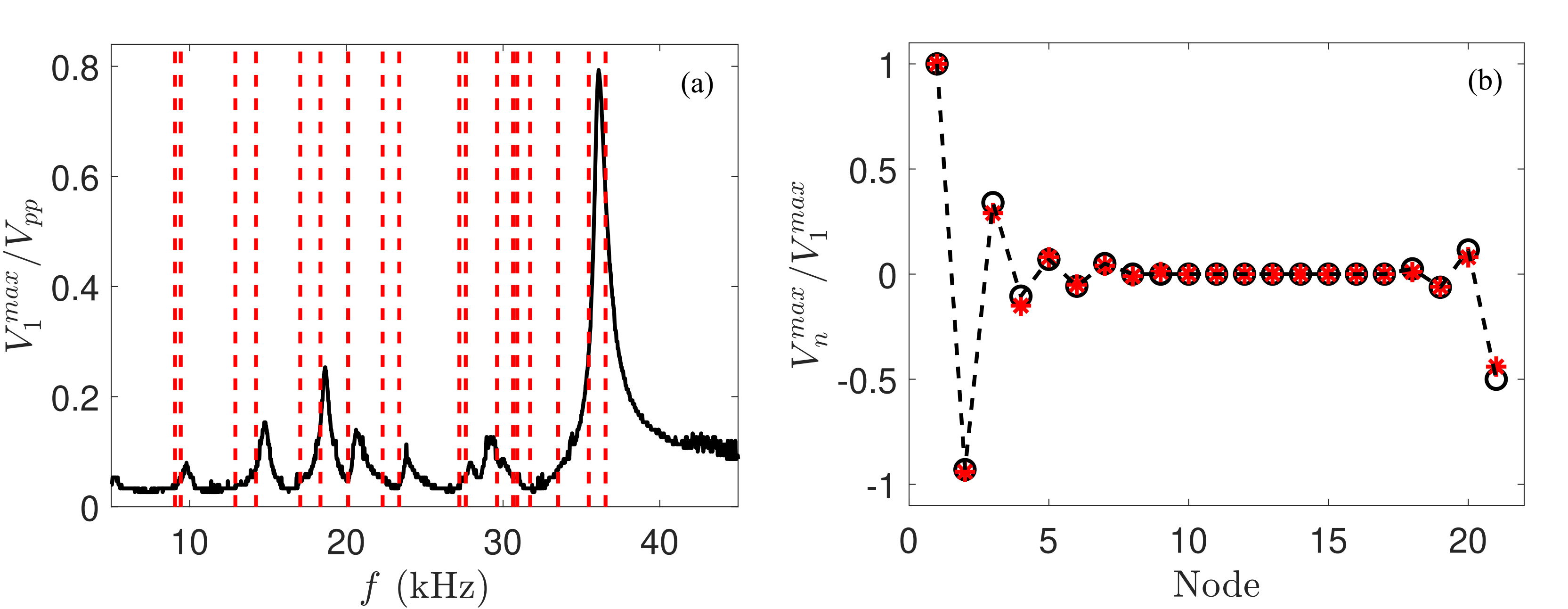}
\caption{(a) Resonance curve corresponding to the linear lattice with Anderson disorder. The red dashed vertical lines indicate the theoretical normal mode frequencies, while the continuous black curve represents the experimentally measured maximum voltage response. (b) Comparison of the experimental and theoretical voltage profiles of the strongly localized normal mode theoretically predicted at 35.46 kHz and experimentally detected at 36.1 kHz. Black circles represent the experimental measurements, whereas red asterisks indicate the theoretical predictions. The lattice was driven at node $n=1$, with $N=21$, $L=0.1$ H, $C_0=600$ pF, $C_d=10$ pF, and $V_{pp}=3$ V. The additional capacitances are listed in Table~\ref{tab:additional_capacitors}.}
\label{fig4}
\end{figure}

In the experimental lattice, we used inductors with an inductance $L=0.1\mathrm{H}$. The lattice was excited at node $n=1$ through a coupling capacitor $C_{\mathrm{d}}=10$ pF, using a sinusoidal voltage source with a peak-to-peak voltage of $V_{pp}=3$ V. The voltage response was measured at the same node. Figure~\ref{fig4}(a) compares these theoretical frequencies with the experimentally measured resonance curve, the theoretical predictions and experimental results are in very good agreement. Focusing on the strongly localized at node 1 mode theoretically predicted at 35.46 kHz and experimentally detected at 36.1 kHz, we measured the voltage profile across the lattice. Figure~\ref{fig4}(b) compares the theoretical and experimental profiles, which once again exhibit a high degree of agreement, confirming both the location and the strong spatial localization of this mode.

Not all normal modes are visible in the experimental frequency-response curve because the voltage is measured locally at node~1. The contribution of a given mode $m$ to this local response depends on its modal amplitude $V_{(m)}$ at that node. Consequently, modes having a node or a very small amplitude at node~1 produce weak or practically undetectable resonance peaks, even though they exist in the complete spectrum. This effect is particularly important for Anderson-localized modes centered far from the measurement node.  Therefore, the absence of a peak in the response measured at node~1 does not imply the absence of the corresponding normal mode, but only that its local spectral weight is too small to be clearly resolved. In general, the theoretical and experimental curves are in good agreement. The small discrepancies between the theoretical and experimental values can be attributed mainly to component tolerances, parasitic capacitances and resistances, dissipative losses not fully included in the idealized theoretical model, and measurement uncertainties. These effects may slightly shift the resonance frequencies. 

\begin{figure}[htbp]
    \centering
    \includegraphics[width=\textwidth]{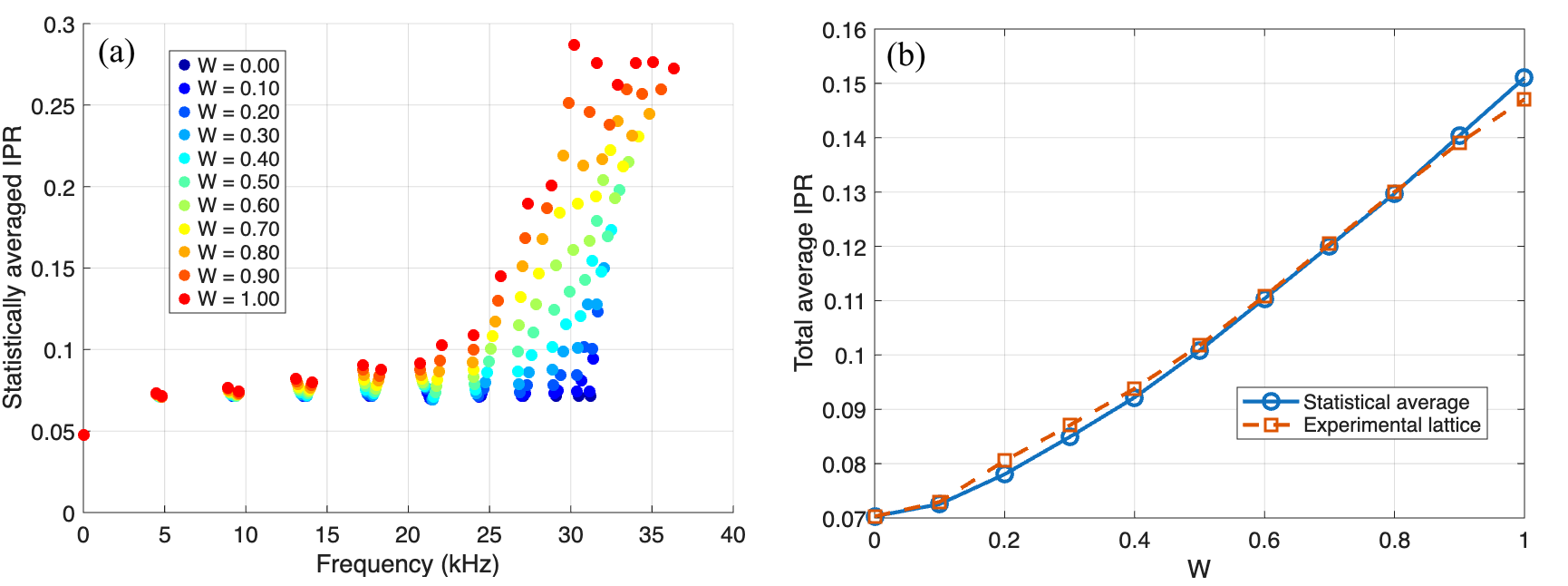}
    \caption{(a) The statistically averaged frequency and IPR for 50 random realizations of $\Delta C_n$ with different disorder strength $W$. (b) Total averaged IPR over all linear modes as a function of $W$. Blue and orange curves correspond to 50 random realizations and the fixed ordering of $\Delta C_n$ used in the experimental lattice, respectively.}
    \label{iprfw}
\end{figure}

We also studied the process of generating Anderson localization by increasing the degree of disorder in the system. A dimensionless parameter $W$ is introduced to control the strength of the capacitive disorder, so that $C_n=C_0+\bar{C_n^{*}}+W\Delta C_n$, where $\Delta C_n=C_n^{*}-\bar{C_n^{*}}$ is the disorder distribution. The values of $C_n^{*}$ were randomly distributed among the lattice nodes. 
For each disorder strength $W$, we perform 50 random realizations to compute $\Delta C_n$. The statistically averaged frequency and IPR are then calculated over these realizations as Fig.~\ref{iprfw}(a) shows. For each $W$, the 21 points represent the linear modes ordered by increasing frequency. 
The results show that the modes become progressively more localized as $W$ increases, reflecting a crossover from weakly localized modes that appear extended in finite-size lattices to strongly localized Anderson modes.
As shown in Fig.~\ref{iprfw}(b), we also compared the total average IPR for the fixed ordering of $\Delta C_n$ used in the experiment with that for 50 random realizations of the ordering of $\Delta C_n$. The arrangement of $\Delta C_n$ chosen here shows no particular significance. Both curves exhibit a consistent trend, with the IPR increasing monotonically as $W$ grows.

\subsection{The DBs in nonlinear chain with Anderson disorder}

\begin{figure}[h]
\centering
\includegraphics[width=0.48\textwidth]{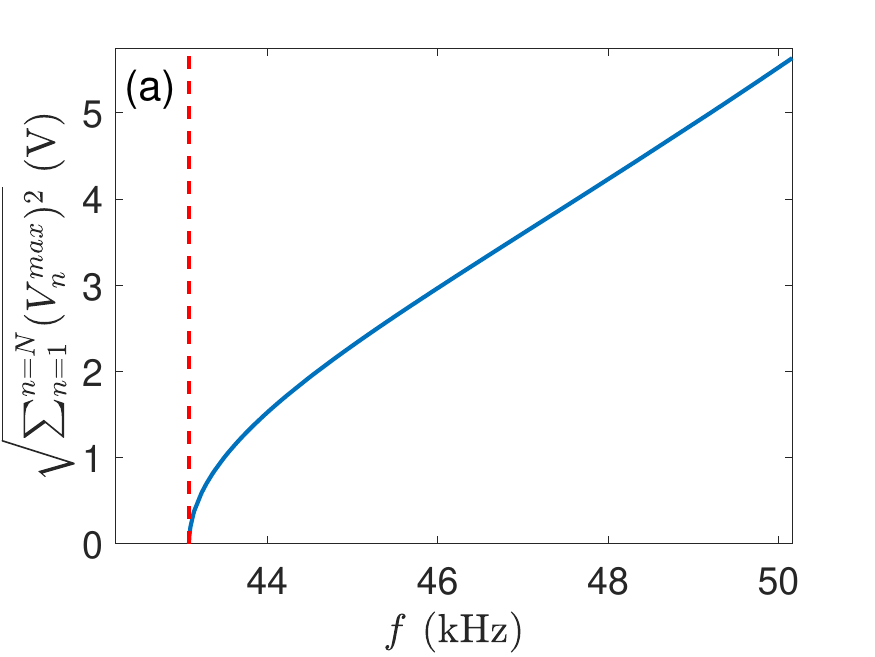}
\includegraphics[width=0.48\textwidth]{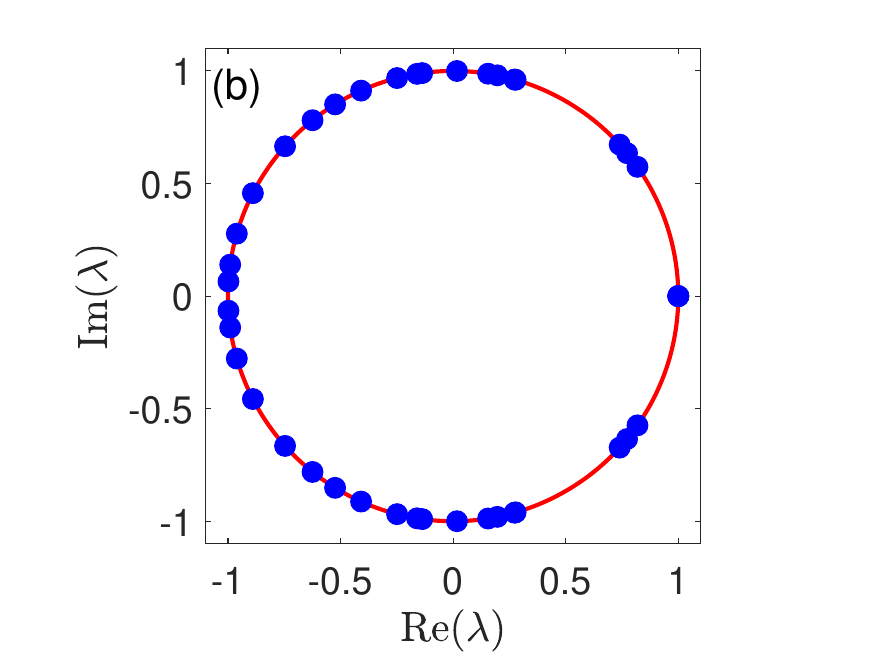}
\includegraphics[width=0.48\textwidth]{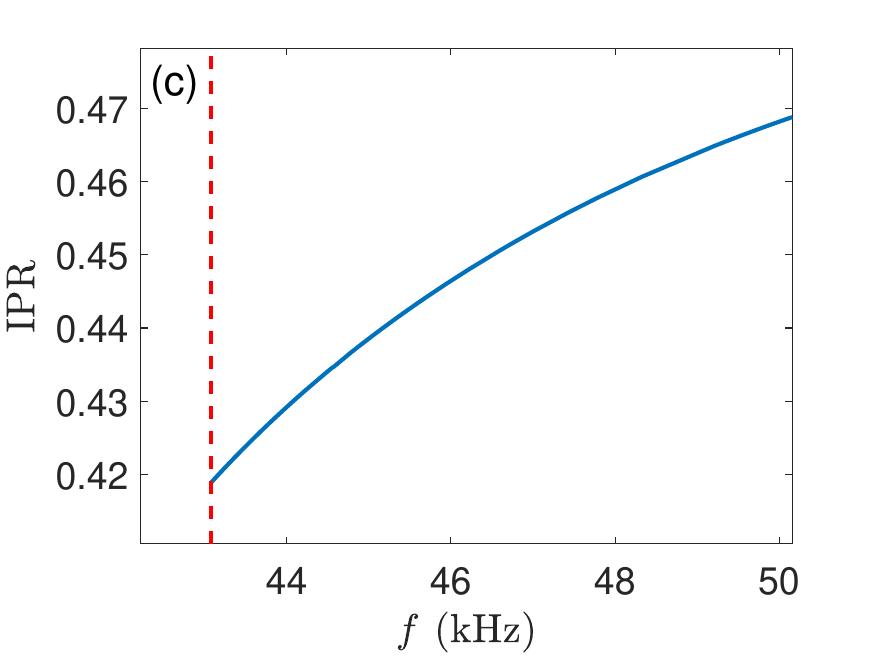}
\includegraphics[width=0.48\textwidth]{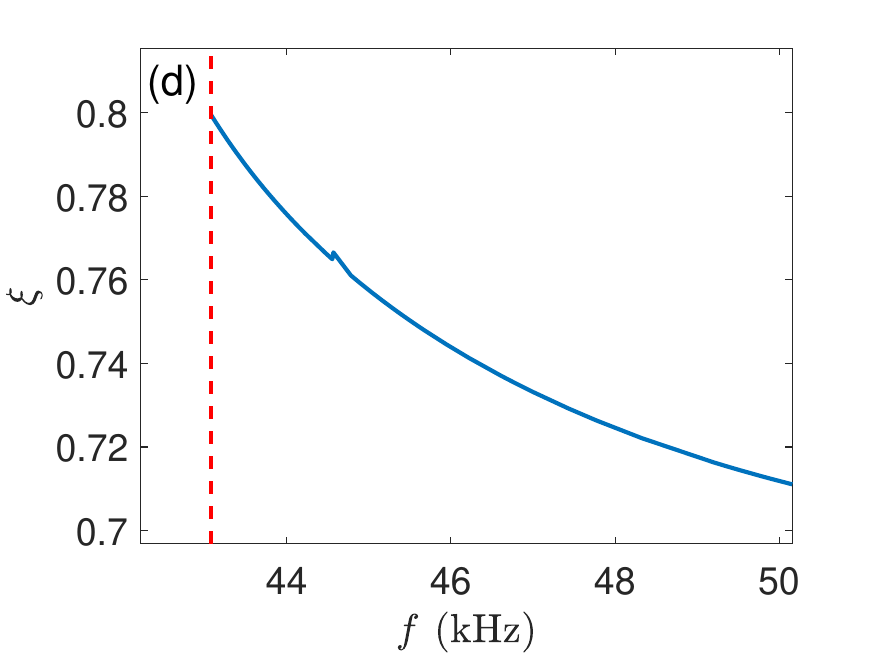}
\caption{Summary of the numerical results for the family of nonlinear periodic solutions of DBs emerging from the linear-limit mode at f=43.08 kHz. (a) $\sqrt{\sum_{n=1}^{n=N} (V_n^{max})^2}$ as a function of frequency. (b) Representative Floquet multiplier spectrum along the solution branch. (c) Inverse participation ratio IPR and (d) exponential localization length, $\xi$, as functions of frequency. The dashed red line indicates the frequency of the corresponding linear mode.}
\label{fig7}
\end{figure}

For the nonlinear model, in the linear limit, Eq.~(\ref{And_eq_14}) reduces the system to the case considered previously, so all its normal modes can be readily determined. The normal-mode frequencies, IPR values, and principal localization nodes are listed in Table~\ref{tab:modes}.
\begin{table}[htbp]
\centering
\caption{Nonlinear Modes and IPR}
\label{tab:modes}
\renewcommand{\arraystretch}{1.1} 
\small 
\begin{minipage}[t]{0.48\textwidth} 
    \centering                        
    \begin{tabular}{c c c c}
        \toprule
        {\textbf{Mode}} & {\textbf{\(f\) (kHz)}} & {\textbf{IPR}} & {\textbf{Main node}} \\
        \midrule
        0 & 0.00 & 0.048 & all \\
        1 & 4.86 & 0.070 & 8 \\
        2 & 5.74 & 0.070 & 3 \\
        3 & 10.20 & 0.08 & 20 \\
        4 & 10.80 & 0.08 & 8 \\
        5 & 14.51 & 0.10 & 14 \\
        6 & 16.40 & 0.09 & 5 \\
        7 & 19.20 & 0.09 & 13 \\
        8 & 20.97 & 0.08 & 1 \\
        9 & 22.59 & 0.11 & 1 \\
        10 & 25.58 & 0.17 & 13 \\
        \bottomrule
    \end{tabular}
\end{minipage}                        
\hfill 
\begin{minipage}[t]{0.48\textwidth}
    \centering
    \begin{tabular}{c c c c}
        \toprule
        {\textbf{Mode}} & {\textbf{\(f\) (kHz)}} & {\textbf{IPR}} & {\textbf{Main node}} \\
        \midrule
        11 & 26.42 & 0.13 & 4 \\
        12 & 32.15 & 0.11 & 16 \\
        13 & 32.33 & 0.11 & 16 \\
        14 & 35.00 & 0.16 & 4 \\
        15 & 36.46 & 0.54 & 9 (19) \\
        16 & 36.64 & 0.66 & 19 (9) \\
        17 & 37.76 & 0.26 & 12 \\
        18 & 39.78 & 0.41 & 7 \\
        19 & 43.08 & 0.42 & 1 \\
        20 & 44.67 & 0.33 & 13 \\
        \bottomrule
    \end{tabular}
\end{minipage}
\end{table}
We focus on the mode with f=43.08 kHz, which is the counterpart of the 35.46 kHz mode in the previous configuration, with a similar profile. This mode is strongly localized around node $n$=1, with IPR=0.42. The results of the numerical continuation and Floquet stability analysis of DBs are presented in Fig.~\ref{fig7}. A stable branch of nonlinear periodic solutions emerges from the linear limit, along which both the oscillation frequency and amplitude increase monotonically. Although the continuation was terminated at the maximum frequency shown in the figure, no limiting point was observed, and the branch can be continued to considerably higher frequencies. Throughout the entire frequency interval investigated, the periodic solutions of DBs remain stable. 

\begin{figure}[h]
\centering
\includegraphics[width=0.6\textwidth]{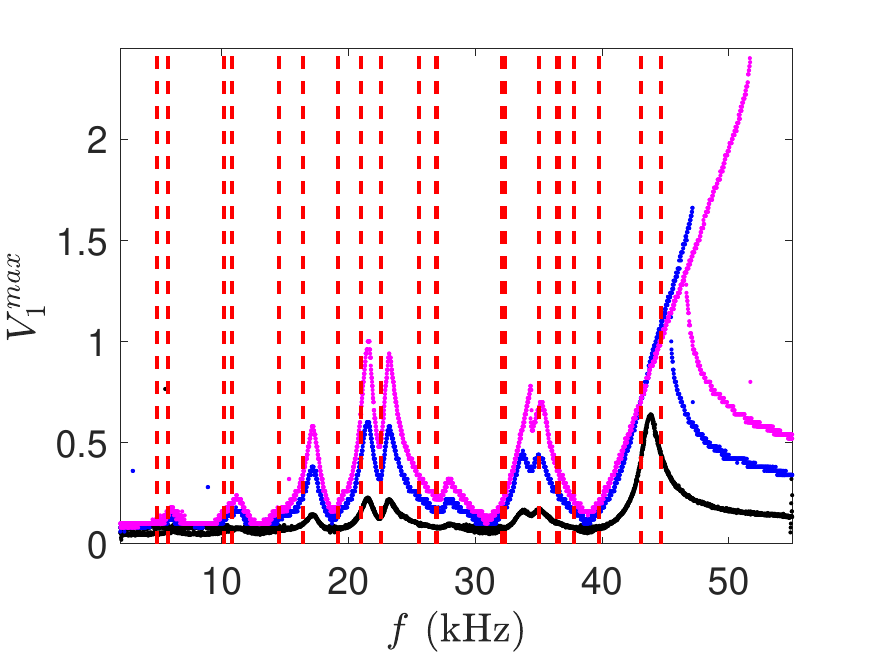}
\caption{Resonance curves for the nonlinear lattice with Anderson disorder. The red dashed vertical lines indicate the theoretically predicted linear normal-mode frequencies, whereas the solid curves represent the experimentally measured maximum voltage response at the driven node $n$=1. The excitation amplitudes are $V_{\mathrm{pp}}=1$ V (black, nearly linear regime), $V_{\mathrm{pp}}=3 $ V (blue, moderately nonlinear regime), and $V_{\mathrm{pp}}=5$ V (magenta, strongly nonlinear regime). The remaining circuit parameters are $N=21$, $L$=0.1 H, and $C_d$=10 pF.}
\label{fig8}
\end{figure}

Returning to the experimental lattice, we analyzed its frequency response, as shown in Fig.~\ref{fig8}. In the nearly linear regime, the experimentally observed resonances agree well with the theoretically predicted normal-mode frequencies. As the strength of the nonlinearity increases, the response extends over a progressively broader frequency interval as the amplitude of the external excitation is increased. This behavior is expected because stronger driving excites higher-amplitude periodic orbits belonging to the nonlinear continuation of the linear modes. The effect is particularly clear for the mode theoretically predicted at 43.08 kHz, which is strongly localized around node $n=1$, where the lattice is both driven and measured. The corresponding curves exhibit the characteristic bending of a nonlinear resonance branch, together with the associated hysteresis loop. Once again, not all resonances are visible in the measured response. Modes localized far from node $n=1$ have a very small amplitude at the excitation and measurement site and therefore produce extremely weak peaks that are effectively undetectable in practice.

\begin{figure}[h]
\centering
\includegraphics[width=0.48\textwidth]{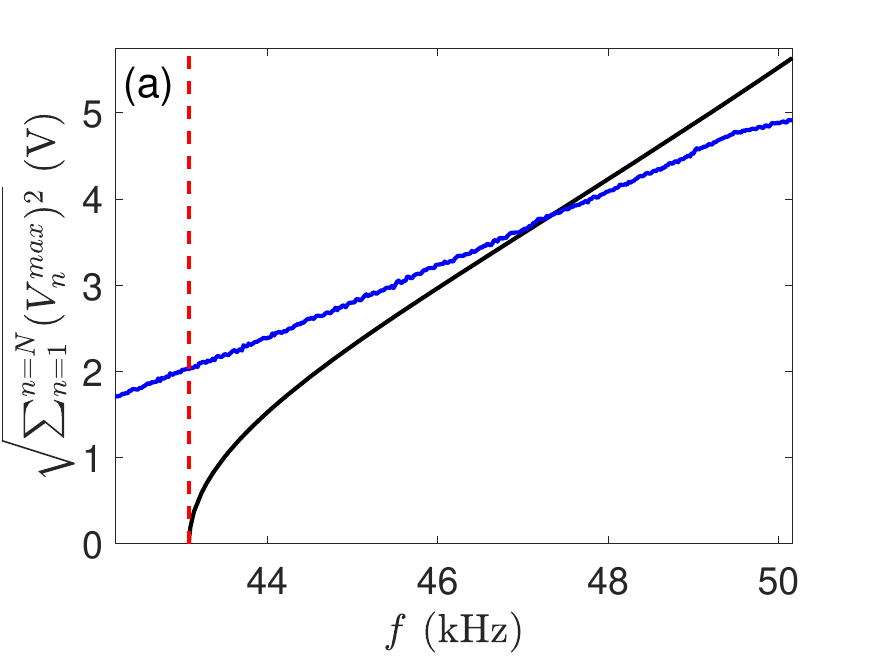}
\includegraphics[width=0.48\textwidth]{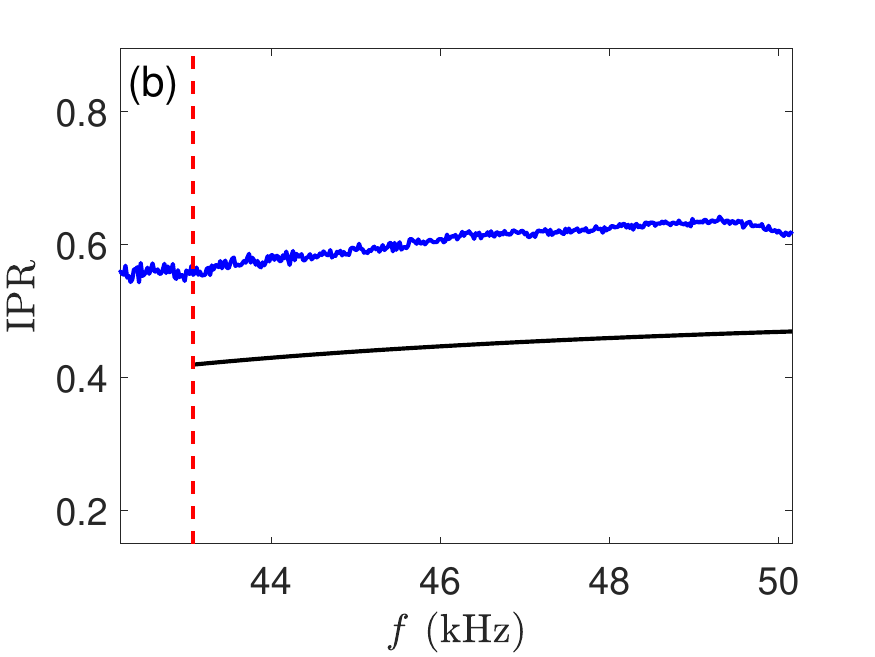}
\caption{Summary of the numerical (black)  and experimental (blue) results for the family of nonlinear periodic solutions emerging from the linear-limit mode at f=43.08 kHz. (a) $\sqrt{\sum_{n=1}^{n=N} (V_n^{max})^2}$ as a function of frequency and (b) Inverse participation ratio IPR. The dashed red line indicates the frequency of the corresponding linear mode.}
\label{fig9}
\end{figure}

\begin{figure}[h]
\centering
\includegraphics[width=0.48\textwidth]{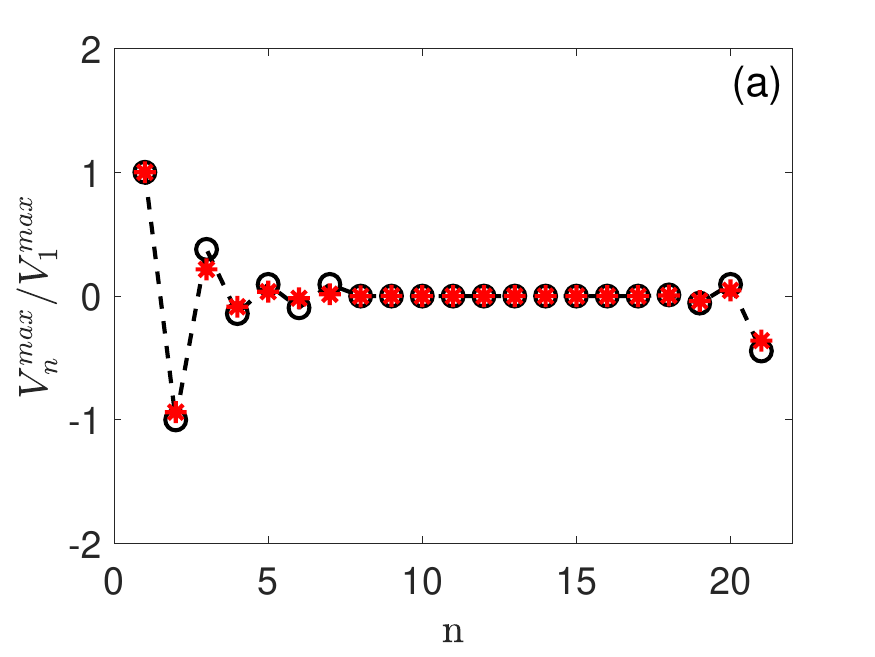}
\includegraphics[width=0.48\textwidth]{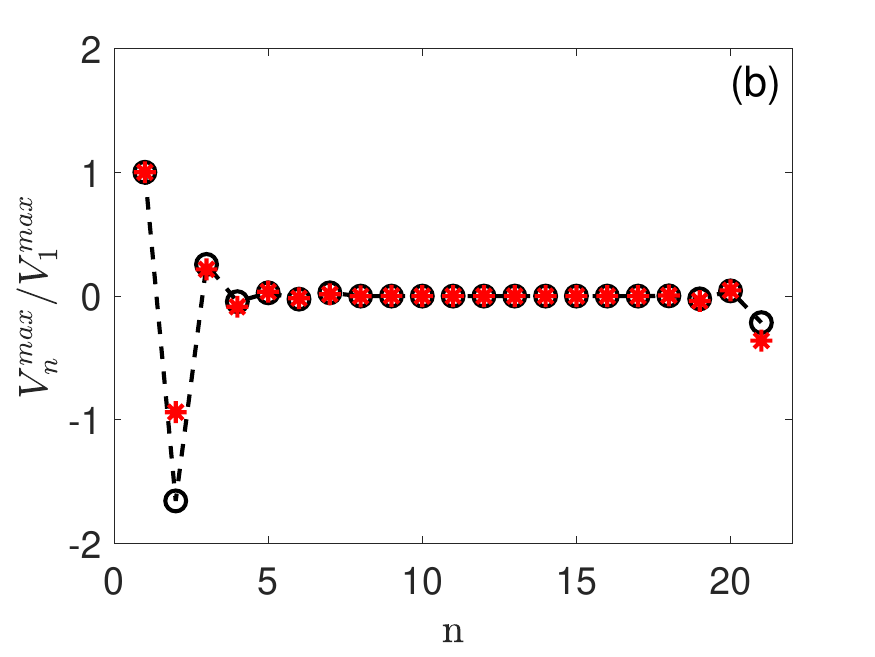}
\caption{Comparison of the experimental and theoretical voltage profiles of the nonlinear localized solutions of DBs emerging from the strongly localized linear normal mode predicted at $f=43.08,\mathrm{kHz}$. Black circles represent the experimental measurements, whereas red asterisks indicate the theoretical predictions. The lattice was driven at node $n=1$, with $N=21$, $L=0.1,\mathrm{H}$, and $C_d=10 \ \mathrm{pF}$. The profiles correspond to (a) $f=43.8 \ \mathrm{kHz}$ and (b) $f=50 \ \mathrm{kHz}$.}
\label{fig10}
\end{figure}

Figure~\ref{fig9} shows the quantities $\sqrt{\sum_{n=1}^{N}\left(V_n^{\max}\right)^2}$ and the IPR obtained from both the theoretical calculations presented above and the experimental measurements. Although some quantitative discrepancies are observed, mainly due to component tolerances and the unavoidable inhomogeneities of the experimental lattice, the theoretical and experimental results exhibit the same overall behavior, as expected. Regarding the spatial structure of the mode, Fig.~\ref{fig10} shows the voltage profiles of DBs, we found that the voltage amplitude at node $n=$2 increases along the branch. This redistribution is accompanied by an increase in the IPR, indicating that the overall mode profile becomes more spatially localized.

\begin{figure}[htbp]
    \centering
    \includegraphics[width=\textwidth]{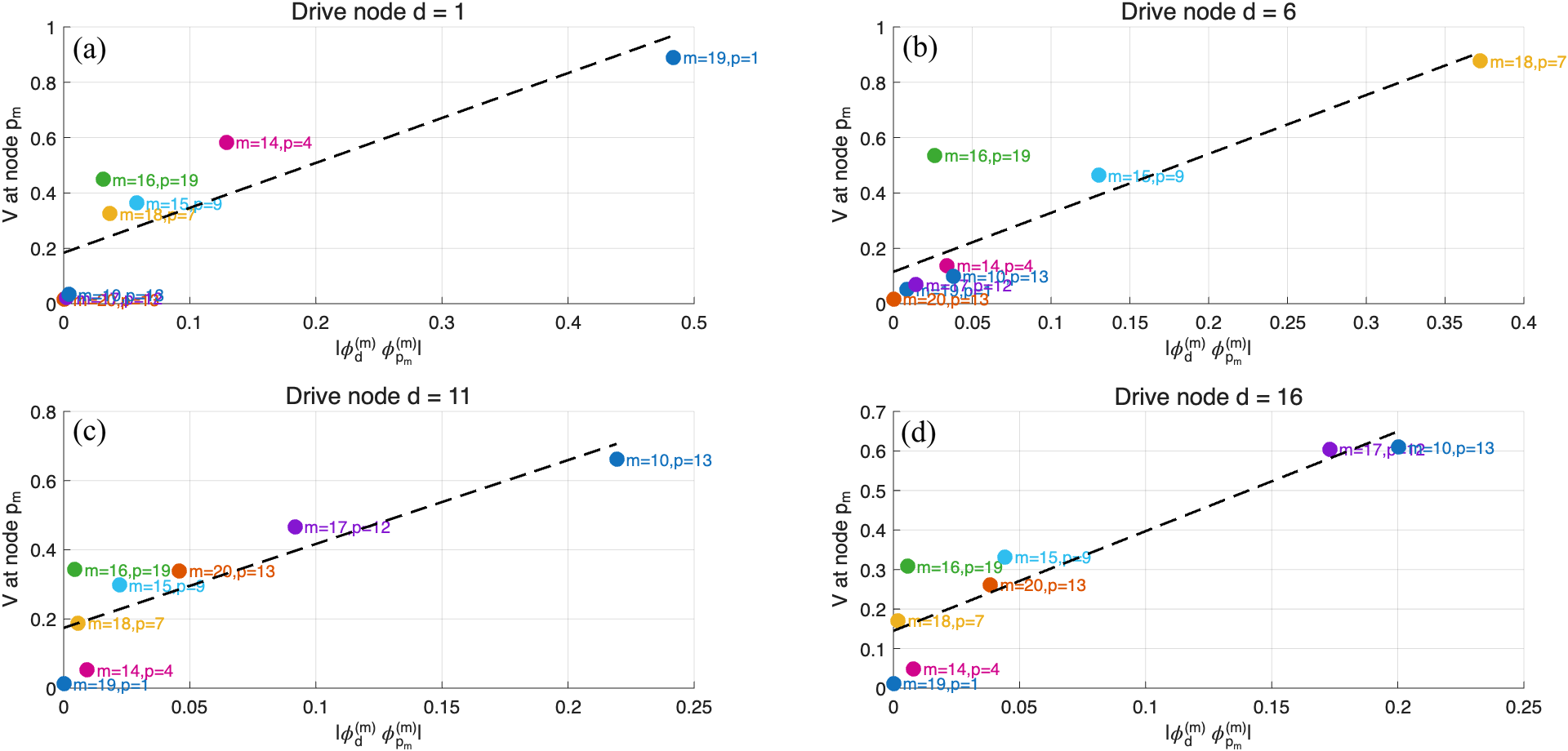}
    \caption{The theoretical result of the relationship between the resonant voltage response of the main nodes for the selected modes $m$ and $|\phi_d^{(m)}\phi_p^{(m)}|$ when drive node (a) $d=1$, (b) $d=6$, (c) $d=11$ and (d) $d=16$. The remaining circuit parameters are $N=21$, $L$=0.1 H, $C_d$=10 pF, $g=0.1$, and $V_d=0.3$V.}
    \label{rs}
\end{figure}

In the case we discussed above, the lattice was driven and measured
at node $n=1$. Consequently, modes localized far from this node produced
very weak or undetectable resonance peaks. We do the theoretical exciting the lattice at different nodes $n=1$, $6$, $11$, and $16$. Then we choose some modes to observe, we use $m$ to represent the mode number.
Because the system exhibits strong nonlinearity, the resonance peak undergoes a significant frequency shift. Therefore, we retained the frequency-sweep procedure to search for the maximum target-node voltage within a $\pm0.05$kHz interval around the target frequency, so as to ensure that the measured resonance strength is closer to the true value for this mode. Although some modes did not reach the maximum target-node voltage within the preset frequency range, we can still basically observe the general trend of the resonance strength. When the system is driven at node $n=d$ and measured at node $n=p$, the relationship between the strength of the $m$th resonance and the product $|\phi_d^{(m)}\phi_p^{(m)}|$ are shown in Fig.~\ref{rs}, where $\phi_n^{(m)}$ denotes the amplitude of the $m$th linear mode at node $n$. In general, the location of the driving point determines which mode receives energy preferentially. A node $p$ closer to node $d$ is more likely to have its resonance peak excited. From the fitted line, the strength of the $m$th resonance is essentially proportional to $|\phi_d^{(m)}\phi_p^{(m)}|$. The coupling strength differs among modes. Some modes absorb energy more readily and thus produce a strong resonance peak, for example the mode with $m=$15, 16. These two modes have two main nodes and the distance between the two main nodes is approximately $N/2$, that makes them more prone to resonance regardless of which point is driven. When one mode is excited, another mode will also be indirectly excited.

\section{CONCLUSIONS AND FUTURE WORK}

In this work, we have investigated linear and nonlinear localization in a disordered electrical lattice composed of $N=21$ coupled LC resonators. 
In the linear regime, the normal-mode spectrum and the corresponding spatial voltage profiles were determined theoretically and compared with experimental measurements. The results confirm the existence of strongly localized Anderson modes produced by the distributed capacitive disorder. In general, good agreement was obtained between theory and experiment in terms of resonance frequencies, localization centers, voltage profiles, IPR values, and localization lengths. The remaining quantitative discrepancies can mainly be attributed to component tolerances, parasitic elements, dissipative losses, and experimental uncertainties. We have also present the result of average IPR theoretically to observe the progressive evolution from extended normal modes to strongly localized Anderson modes.

Nonlinearity was introduced by replacing the linear capacitors with nonlinear elements based on NTE618 varactor diodes. The localized linear mode at $f=43.08 \ \mathrm{kHz}$ was then continued into a family of nonlinear periodic solutions of DBs. The numerical continuation shows that both the frequency and the amplitude increase along the branch. The corresponding Floquet analysis indicates that the DBs remain spectrally stable throughout the investigated frequency interval, while their IPR and localization length reveal the evolution of their spatial localization. When the system is driven at different nodes, a node which is closer to the driving node is more likely to have its resonance peak excited. If two different modes have two same main nodes, when one node is excited, another node will also be indirectly excited.
The experimental nonlinear response is consistent with these numerical predictions. As the driving amplitude increases, the response associated with the localized mode extends over a broader frequency interval and exhibits the characteristic bending and hysteresis of a nonlinear resonance. Overall, the results demonstrate how linear Anderson-localized modes can evolve into stable, disorder-pinned DBs.

\section{ACKNOWLEDGMENTS}
The authors thank Lars English (Dickinson College)
for his valuable assistance and J. Cuevas-Maraver
(University of Seville) for helpful comments and suggestions.

\end{document}